\documentclass[aps,prb]{revtex4}
\usepackage{amsmath}
\usepackage{graphicx}

\begin{document}
\title{Quantum Vavilov-Cherenkov radiation from a small neutral particle moving parallel to a     transparent dielectric plate}

\author{A.I.Volokitin$^{1,2,3}$\footnote{Corresponding author.
\textit{E-mail address}:alevolokitin@yandex.ru}    and B.N.J.Persson$^1$}
 \affiliation{$^1$Peter Gr\"unberg Institut,
Forschungszentrum J\"ulich, D-52425, Germany}

\affiliation{
$^2$Samara State Technical University, Physical Department, 443100 Samara, Russia}
\affiliation{
$^3$Samara State Aerospace University, Physical Department, 443086 Samara, Russia}

\begin{abstract}

We study the quantum Vavilov-Cherenkov (QVC) radiation and quantum friction occurring during motion of a small neutral particle parallel to a transparent dielectric plate with the refractive index $n$. 
This phenomenon occurs  above the threshold velocity $v_c=c/n$. The particle acceleration and rate of heating are determined by the friction force and heating power in the rest reference frame of the particle. We calculate these quantities  starting from the expressions for the friction force and the radiative energy transfer in the  plate-plate configuration, assuming  plate  at the rest in the \textit{lab} frame rarefied. Close to the light velocity there is a big difference between the friction force and the radiation power in the rest frame of a particle and in the \textit{lab} reference frame. This difference is connected with  the change of the rest mass of the particle  due to absorption of radiation. Close to the threshold velocity the decrease of the kinetic energy  of the particle is determined mainly by radiation power  in in the \textit{lab} frame. However, close to the light velocity it is determined also by the heating power for the particle.  We establish the connections between the quantities in the different reference frames.  For a nanoparticle the QVC radiation intensity can be comparable to  classical one. We discuss the possibility to detect QVC radiation. 
\end{abstract}
\maketitle


\vskip 5mm

\section{Introduction}
A remarkable manifestation of the interaction between the electromagnetic field and matter is the
emission of light by charged particle  moving at a constant  superluminally velocity in a medium: the Vavilov-
Cherenkov (VC)  radiation \cite{Cherenkov1934,Vavilov1934,Cherenkov1937,Tamm1937,Tamm1959,Ginzburg1996}, which has broad applications in the detection
of high-energy charged particles in astrophysics and particle physics. This phenomenon can be explained by classical electrodynamics because moving charge
corresponds to a time-dependent current. The problem of the interaction of a point charge
with the electromagnetic field in a medium can be reduced to the solution of the equation of the motion  for  harmonic
oscillators under the action of  external forces with frequencies
\begin{equation}
\omega=\mathbf{q\cdot v}=qv\mathrm{cos}\theta=\frac{\omega(q)v\mathrm{cos}\theta}{v_0},
\end{equation}
where $\theta$ is the angle between the wave vector $\mathbf{q}$ and velocity $\mathbf{v}$,
$v_0=c/n$ is the phase velocity of the light in the medium with the refractive index $n$, $\omega(q)=v_0q$ is
the eigenfrequency of the electromagnetic wave in the medium. At $\omega=\omega(q)$, there is resonance,
and the amplitudes of oscillators grow with time, i.e. radiation occurs. At resonance
\begin{equation}
\mathrm{cos}\theta=\frac{v_0}{v}<1.
\end{equation}
Thus radiation occurs only if the charge velocity $v$ exceeds the phase velocity of light
in a medium:  $v>v_0$.

A neutral particle  moving at a constant velocity can also radiate due to presence of the fluctuating dipole moment.
This radiation was  described by  Frank \cite{Frank1943} and   Ginzburg and  Frank \cite{Ginzburg1945a,Ginzburg1945b} (see also \cite{Tamm1959,Ginzburg1996} for
review of these work). If a particle has no internal degrees of freedom (e.g., a point charge), then the energy of the radiation is determined by the change of
the  kinetic energy of the object. However, if a particle has  internal degrees of freedom (say, an atom), then two types  of radiation
are possible.  If the frequency of the radiation in the \textit{lab} reference frame $\omega >0$, then in  the rest frame of a particle, due to the Doppler effect
the frequency of the radiation $\omega^{\prime}=\gamma(\omega-q_xv)$, where $\gamma = \sqrt{1-\beta^2}$, $\beta=v/c$, $q_x$ is the projection of the wave vector on the direction of the motion. In the normal Doppler effect region,
when $\omega^{\prime}>0$, the radiated energy is determined by the decrease of the internal energy. For example, for an atom the state changes from the excited state $|1>$ to the
ground state
$|0>$ in which case $\omega^{\prime}=\omega_0$ where $\omega_0$ is the transition frequency. The region of the \textit{anomalous} Doppler effect corresponds to $\omega^{\prime}<0$ in which case a particle becomes excited when it radiates. For example, an atom could
 experience the transition from the ground state $|0>$  to the excited state $|1>$ when it radiates in which case $\omega^{\prime}=-\omega_0$.
In such a case energy conservation requires that the energy  of the radiation and of the excitation
result from a
decrease of the kinetic energy of the particle. That is, the self-excitation of a system is accompanied by a slowing down of the motion of the particle as a whole. If the particle
is moving relative to the transparent dielectric medium  the propagating electromagnetic  waves can be excited in the medium in which
case $\omega=v_0q$. Thus the condition for the occurrence of the radiation has the form
\begin{equation}
q_xv=v_0q+\frac{\omega_0}{\gamma}>v_0q_x+\frac{\omega_0}{\gamma}.
\end{equation}
It follows that  radiation occurs only when $v>v_c=v_0$, which coincide for condition of the radiation from a point charge. Due to the quantum origin, the radiation produced by a neutral
particle is denoted as the quantum Vavilov-Cherenkov (QVC) radiation, and the friction associated with this radiation is denoted as quantum friction, which is the limiting case of the
Casimir friction  produced by thermal and quantum fluctuations of the electromagnetic field; the
same fluctuations also give rise to the Casimir -- van der Waals forces \cite{Casimir1948,Lifshitz1955,Dalvit2011}. The link between QVC radiation and quantum friction in the plate-plate configuration was  discussed  within framework of  a non-relativistic    \cite{KardarPRA2013}, semi-relativistic \cite{PendryJMO1998}  and fully relativistic theories  \cite{VolokitinPRB2016}. 

The thermal and quantum fluctuation of the current density in one body induces a current density
in other body; the interaction between these current densities is the origin of the
Casimir interaction. When two bodies are in relative motion, the induced current will
lag slightly behind the fluctuating current inducing it, and this is the origin of the
Casimir friction. At present the radiation friction is attracting a lot of attention due to the fact that it is one of the mechanisms of noncontact
friction between bodies in the absence of direct contact \cite{VolokitinRMP2007}. The noncontact friction determines the ultimate limit to which the friction force can be reduced and,
consequently, also the force fluctuations because they are linked to friction via  the fluctuation-dissipation theorem. The force fluctuations (and hence friction) are important for ultrasensitive force detection.

Radiative friction has deep roots in the foundation of quantum mechanics. The friction force acting on  a particle moving relative to the blackbody radiation, which is a
limiting case of the fluctuating electromagnetic field, was studied by Einstein and Hopf \cite{Einstein1910} and Einstein \cite{Einstein1917} in the early days of quantum mechanics.
The friction induced by the electromagnetic fluctuations has been studied in the configurations plate-plate \cite{PendryJPCM1997,PendryJMO1998,VolokitinJPCM1999,VolokitinPRL2003,VolokitinPRB2003,VolokitinRMP2007,VolokitinPRB2008,KardarPRA2013,VolokitinPRB2016},  neutral particle-plate \cite{TomassonePRB1997,VolokitinPRB2002,VolokitinRMP2007,DedkovPLA2005,DedkovJPCM2008,BrevikEntropy2013,KardarPRD2013,BrevikEPJD2014,BartonNJP2010,HenkelNJP2013,
DalvitPRA2014,VolokitinNJP2014,HenkelJPCM2015}, and neutral particle-blackbody radiation  \cite{VolokitinRMP2007,VolokitinPRB2008,HenkelNJP2013,MkrtchianPRL2003,DedkovNIMPR2010,VolokitinPRA2015}.
While the predictions of the theory for the Casimir forces were verified in many experiments \cite{Dalvit2011}, the detection of the
Casimir friction  is still challenging problem for  experimentalists. However, the frictional  drag between quantum wells \cite{GramilaPRL1991,SivanPRL1992} and  graphene sheets
\cite{KimPRB2011,GeimNaturePhys2012}, and the current-voltage dependence of nonsuspended graphene on the surface of the polar dielectric SiO$_2$ \cite{FreitagNanoLett2009}, were accurately described using the
theory of the Casimir friction \cite{VolokitinPRL2011,VolokitinJPCM2001b,VolokitinEPL2013}.

It was shown in  Ref. \cite{VolokitinNJP2014}, that the friction force on a small neutral particle
at relativistic motion parallel to  flat surface of a
dielectric in the \textit{lab} frame can be obtained from the friction force between two plates assuming
that the moving  plate is sufficiently rarified and consists of particles with  the
electric polarizability  $\alpha$. In this paper,  we extend this approach to calculate the friction force in the rest frame of the particle.

\section{Theory}

\begin{figure}[tbp]
\includegraphics[width=0.80\textwidth]{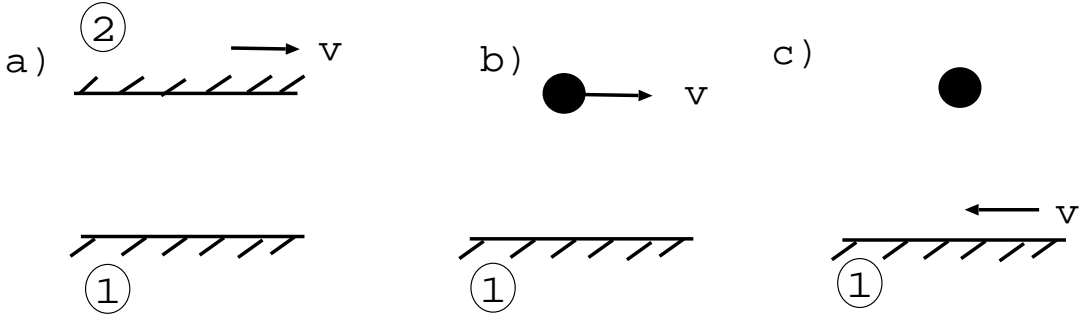}
\caption{The schemes of the configurations (a) plate-plate, (b) particle-plate in the \textit{lab} frame and (c) particle-plate in the rest frame of a particle. The friction force in the  particle-plate configuration    can be obtained from the friction force in the  plate-plate configuration  assuming rarefied the plate 2.
\label{Fig1}}
\end{figure}

 We consider two plates having flat parallel surfaces separated
by a distance $d$ and moving with the velocity $v$ relative to each other,
see Fig. \ref{Fig1}. We introduce the two reference frames $K$ and $K^{\prime }$ which are the rest reference frames for the plates 1 and
2, respectively. According to a fully relativistic theory     \cite{VolokitinPRB2008}    the friction force  at zero temperature
$F_{1x}$ (denoted  as quantum friction \cite{PendryJPCM1997}) acting on the surface of plate 1, and the radiation power  $W_1$ absorbed by plate  1 in the    $K$ frame, are determined by formulas
\begin{equation}
\left(\begin{array}{c}
F_{1x}\\
W_1
\end{array} \right)
 = \int_{-\infty}^{\infty} \frac{dq_y}{2\pi}\int_{0}^{\infty} \frac{dq_x}{2\pi}
\int_0^{q_xv} \frac{d\omega}{2\pi} \left(\begin{array}{c}
\hbar q_x\\
\hbar \omega
\end{array} \right)\Gamma_{12}(\omega, \mathbf{q}),
\label{qvc3}
\end{equation}
where the positive quantity
\[
\Gamma_{12}(\omega, \mathbf{q})= - \frac{4}{|\Delta|^2}[(q^2 - \beta kq_x)^2 -
\beta^2k_z^2q_y^2]
\{\mathrm{Im}R_{1p}[(q^2 - \beta kq_x)^2\mathrm{Im}R_{2p}^{\prime}|\Delta_{ss}|^2
\]
\begin{equation}
+\beta^2k_z^2q_y^2\mathrm{Im}R_{2s}^{\prime}|\Delta_{sp}|^2]
 +(p\leftrightarrow s)\}e^{-2 k_z d}
\end{equation}
can be identified as a spectrally resolved photon emission rate,
\begin{equation}
\Delta = (q^2 - \beta kq_x)^2\Delta_{ss}\Delta_{pp} - \beta^2k_z^2q_y^2\Delta_{ps}\Delta_{sp},\,\,
\Delta_{pp} = 1 - e^{-2k_zd}R_{1p}R_{2p}^{\prime},\, \Delta_{sp} =
1 + e^{-2k_zd}R_{1s}R_{2p}^{\prime},
\end{equation}
$k_z=\sqrt{q^2-(\omega/c)^2}$, $k=\omega/c$, 
$R_{1 p(s)}$ is the reflection amplitude for surface
1 in the  $K$ frame for   a $p(s)$ - polarized electromagnetic wave,
$R_{2 p(s)}^{\prime} = R_{2 p(s)}(\omega^{\prime}, q^{\prime})$ is the reflection
amplitude for surface
2  in the $K^{\prime}$ frame for   a $p(s)$ - polarized electromagnetic wave,
$\omega^{\prime}=\gamma(\omega-q_xv)$, $q_x^{\prime}=\gamma (q_x-
\beta k)$, $\Delta_{ss}=\Delta_{pp}(p\leftrightarrow s)$, $\Delta_{ps}=\Delta_{sp}(p\leftrightarrow s)$.
The symbol
$(p\leftrightarrow s$) denotes the terms that are obtained from the preceding terms by
exchange of indexes   $p$ and $s$.

Assuming that   the dielectric permittivity
of the rarefied plate is close to the
unity, i.e. $\varepsilon -1\rightarrow 4\pi \alpha N\ll 1$, where $N$ is the
concentration of particles in a plate  in the co-moving reference frame, then 
 to
linear order in the concentration $N$ the reflection amplitudes for the rarefied plate  in the co-moving frame
are
\begin{equation}
R_{p} = \frac{\varepsilon k_z - \sqrt{k_z^2 - (\varepsilon -1)(\omega/c)^2}} {\varepsilon k_z +
\sqrt{k_z^2 - (\varepsilon -1) (\omega/c)^2}}
\approx  N\pi\frac{q^2+k_z^2}{k_z^2} \alpha,\,\,R_{s} = \frac{k_z - \sqrt{k_z^2 - (\varepsilon -1) (\omega/c)^2}}{k_z +\sqrt{k_z^2 - (\varepsilon -1) (\omega/c)^2}}
\approx  N\pi\frac{q^2-k_z^2}{k_z^2} \alpha.
\end{equation}
Because $R_{p(s)}\ll 1$ for the rarefied plate,  it is possible to neglect  the multiple-scattering of the electromagnetic waves 
 between the surfaces.  In this approximation
$\Delta_{pp}\approx \Delta_{ss}\approx \Delta_{sp} \approx \Delta_{sp}\approx 1$,
\begin{equation}
\Delta \approx (q^2 - \beta kq_x)^2 - \beta^2k_z^2q_y^2= \frac{(qq^{\prime})^2}{\gamma^2},
\end{equation}
\begin{equation}
(q^2 - \beta kq_x)^2\mathrm{Im}R_{2p}^{\prime}|\Delta_{ss}|^2
+\beta^2k_z^2q_y^2\mathrm{Im}R_{2s}^{\prime}|\Delta_{sp}|^2]\approx\frac{(qq^{\prime})^2}{\gamma^2} \mathrm{Im}R_{2p}^{\prime}+
\beta^2k_z^2q_y^2\mathrm{Im}(R_{2p}^{\prime}+R_{2s}^{\prime}),
\end{equation}
\[
\Gamma_{12}=-4
\left[\left(\mathrm{Im}R_{1p}\mathrm{Im}R_{2p}^{\prime}+\mathrm{Im}R_{1s}\mathrm{Im}R_{2s}^{\prime}\right)\left(1+\gamma^2\beta^2\frac {k_z^2q_y^2}{q^2q^{\prime 2}}\right)\right.
\]
\begin{equation}
\left.+\gamma^2\beta^2\frac {k_z^2q_y^2}{q^2q^{\prime 2}}\left(\mathrm{Im}R_{1p}\mathrm{Im}R_{2s}^{\prime}+\mathrm{Im}R_{1s}\mathrm{Im}R_{2p}^{\prime}\right)\right].
\label{gamma}
\end{equation}

The friction force $f_x$ acting on a particle, and the radiation power $w$ absorbed by it, can be obtained in the $K$ frame from  
 Eq. (\ref{gamma})  assuming
the plate 2 as sufficiently rarefied \cite{VolokitinNJP2014} (see Fig. \ref{Fig1}b). In this case the friction force acting on the surface 2,   $F_{2x}$, and the radiation power absorbed by it,  $W_2$, are
\begin{equation}
\left(\begin{array}{c}
F_{2x}\\
W_2
\end{array} \right)
 = \left(\begin{array}{c}
-F_{1x}\\
-W_1
\end{array} \right)=
N^{\prime}\int_d^{\infty} dz \left(\begin{array}{c}
f_{x}(z)\\
w(z)
\end{array} \right),
\label{new1}
\end{equation}
where $N^{\prime}=
\gamma N$  is the concentration of  particles in the plate 2 in the $K$ frame,   
\begin{equation}
\left(\begin{array}{c}
f_{x}(z)\\
w(z)
\end{array} \right)
= \frac{1}{\gamma\pi^2}\int d^2q
\int_0^{vq_x} d\omega \left(\begin{array}{c}
\hbar q_x\\
\hbar \omega
\end{array} \right)\frac{e^{-2 k_z z}}{k_z} [\mathrm{Im}R_{1p}(\omega)\phi_p +
\mathrm{Im}R_{1s}(\omega)\phi_s]\mathrm{Im}\alpha(\omega^{\prime}),
\label{qvc9}
\end{equation}
\[
\phi_p=(\omega^{\prime}/c)^2+2\gamma^2(q^2-\beta^2q_x^2)\frac{k_z^2}{q^2},\,\,
\phi_s=(\omega{^\prime}/c)^2+2\gamma^2\beta^2q_y^2\frac{k_z^2}{q^2}.
\]
In the rest reference frame of an object the radiation power absorbed by it is equal to the heating power for the object. Thus $-w$ is equal to the heating power for the plate 1. Eqs. (\ref{qvc9}) agree with the results obtained in \cite{DedkovJPCM2008,HenkelNJP2013,VolokitinNJP2014}.
However, as  shown in Refs.  \cite{VolokitinPRB2008,VolokitinPRA2015},
the acceleration and heating of the particle are determined by the friction force
$f_x^{\prime}$  and by the radiation power $w^{\prime}$ absorbed  by the
particle in the  rest reference frame of a particle (the $K^{\prime}$ frame)
\begin{equation}
m_0\gamma^3\frac{dv}{dt}=m_0\frac{dv^{\prime}}{dt^{\prime}} = f_x^{\prime},
\end{equation}
\begin{equation}
w^{\prime}=\frac{dm_0}{dt^{\prime}}c^2
\end{equation}
where $m_0$ is the rest mass of particle, $v^{\prime}\ll v$ and $t^{\prime}$ are the velocity and time in the $K^{\prime}$, respectively. 
These quantities can be also obtained assuming the plate 2 as sufficiently rarefied (see Fig. \ref{Fig1}c). In this case in the $K^{\prime}$ frame the friction force 
acting on the surface 2, $F_{2x}^{\prime}$  and the radiation power absorbed by it, $W_2^{\prime}$, are
\begin{equation}
\left(\begin{array}{c}
F_{2x}^{\prime}\\
W_2^{\prime}
\end{array} \right)
 = \left(\begin{array}{c}
-\tilde{F}_{1x}\\
\tilde{W}_1
\end{array} \right)=
N\int_d^{\infty} dz \left(\begin{array}{c}
f_{x}^{\prime}(z)\\
w^{\prime}(z)
\end{array} \right),
\label{new2}
\end{equation}
where $\tilde{F}_{1x}$ and $\tilde{W}_1$ are obtained from $F_{1x}$ and $W_1$ after the replacement of the indexes $1\leftrightarrow 2$, 
\begin{equation}
\left(\begin{array}{c}
f_{x}^{\prime}\\
w^{\prime}
\end{array} \right)
= \frac{1}{\pi^2}\int_{0}^{\infty} dq_x\int_{-\infty}^{\infty} dq_y
\int_0^{q_xv} d\omega \left(\begin{array}{c}
\hbar q_x\\
-\hbar \omega
\end{array} \right)\frac{e^{-2 k_z d}}{k_z} [\mathrm{Im}R_{1p}(\omega^{\prime})\phi_p^{\prime}
+ \mathrm{Im}R_{1s}(\omega^{\prime})\phi_s^
{\prime}]\mathrm{Im}\alpha(\omega),
\label{qvc11}
\end{equation}
\[
\phi_p^{\prime}=(\omega/c)^2+2\gamma^2(q^{\prime 2}-\beta^2q_x^{\prime 2})
\frac{k_z^2}{q^{\prime2}},\,\,
\phi_s^{\prime}=(\omega/c)^2+2\gamma^2\beta^2q_y^2\frac{k_z^2}{q^{\prime 2}}.
\]
The relation between the different  quantities in the $K$ and $K^{\prime}$ frames can be found using the Lorentz transformations for
the energy-momentum tensor for a plate 2 according to which
\begin{equation}
F_{2x}=\gamma\left(F_{2x}^{\prime}+v\frac{W_2^{\prime}}{c^2}\right),\,\,W_2=\gamma (W_2^{\prime}+vF_{2x}^{\prime}),
\label{qvc16}
\end{equation}
Using  Eqs. (\ref{new1}) and  (\ref{new2}) gives
\begin{equation}
f_x = f_x^{\prime} + v \frac{w^{\prime}}{c^2},\,\,w = w^{\prime} + vf_x^{\prime}.\label{qvc17}
\end{equation}
These relation also can be found  using the Lorentz transformation for
the energy-momentum  for a particle according to which
\begin{equation}
p_x=\gamma\left(p_x^{\prime}+vm_0\right),\,\,\varepsilon=\gamma (m_0c^2+vp_x^{\prime}),
\end{equation}
where $p_x$ and $\varepsilon$ are the momentum and energy of a particle in the $K$ frame, respectively, and $p_x^{\prime}$ and $m_0c^2$ are the same quantities in the $K^{\prime}$ frame. When the derivative 
of 4-momentum is taken with respect to \textit{lab} time, then the factor $\gamma$ disappears in the right place because $dt=\gamma dt^{\prime}$ and the relations (\ref{qvc17}) are obtained  \cite{VolokitinPRA2015}. From the inverse transformations
\begin{equation}
F_{2x}^{\prime}=\gamma\left(F_{2x}-v\frac{W_2}{c^2}\right),\,\,W_2^{\prime}=\gamma (W_2-vF_{2x}),
\label{qvc16prime}
\end{equation}
follows
\begin{equation}
f_x^{\prime} = \gamma^2(f_x - v \frac{w}{c^2}),\,\,w^{\prime} = \gamma^2(w - vf_x).\label{qvc17prime}
\end{equation}
These relations also can be  obtained as above using the Lorentz transformation for
the energy-momentum  for a particle. We note that in the contrast with the relations (\ref{qvc17}) on the right side of the relations (\ref{qvc17prime}) there is an extra factor $\gamma^2$. This is because the friction force and radiation power are not the 4-vectors. The kinetic energy of a particle in the \textit{lab} frame $\varepsilon_K=\varepsilon-m_0c^2$ where the total energy of a particle in the $K$ frame $\varepsilon=\gamma (m_0c^2+p_x^{\prime}v)$ . The rate of change of the kinetic energy
\begin{equation}
\frac{d\varepsilon_K}{dt}=w^{\prime}+f_x^{\prime}v-\frac{w^{\prime}}{\gamma}=w-\frac{w^{\prime}}{\gamma}=vf_x-\frac{(\gamma-1)w^{\prime}}{\gamma^2}.
\label{qvc18}
\end{equation}
Thus the rate of change of the kinetic energy in the $K$ frame is equal to the friction force power in this frame only when $w^{\prime}=0$.

\section{Results}

For the transparent dielectrics the reflection amplitudes are given by the Fresnel's formulas
\begin{equation}
R_p = \frac{in^2k_z-\sqrt{n^2(\omega/c)^2-q^2}}{in^2k_z+\sqrt{n^2(\omega/c)^2-q^2}},\,R_s = \frac{ik_z-\sqrt{n^2(\omega/c)^2-q^2}}{ik_z+\sqrt{n^2(\omega/c)^2-q^2}}.
\label{amplitude}
\end{equation}
There is no restriction on  the imaginary part of the particle polarizability
in the integration range $0<\omega<v_xv$. Thus the integrand in Eq. (\ref{qvc9}) is
nonzero only in the range  $v_0q_x<\omega<q_xv$, where the imaginary part of the reflection amplitude
is nonzero, thus the critical velocity
$v_c=v_0=c/n$. 

Assuming that the imaginary part of the particle polarizability is determined by the formula
\[
\mathrm{Im}\alpha=R^3\omega_0^2\frac{\omega/\tau}{(\omega^2-\omega_0^2)^2+(\omega/\tau)^2},
\]
where $R$ is the radius of the particle, $\omega_0$ is the plasmon frequency for the particle and $\tau$ is the damping constant, then close to the resonance at $\omega \approx \omega_0$
\[
\mathrm{Im}\alpha=R^3\omega_0\frac{\pi}{2}\delta(\omega-\omega_0),
\]
and from Eq. (\ref{qvc11}) the resonant contributions to the friction force and the heating power close to the threshold velocity in the $K^{\prime}$ frame are dominated by the contributions
from $p$-polirized waves, and are given by (see Appendix \ref{A})
\begin{equation}
f_{xp}^{res\prime }=-\frac{\hbar R^3\omega_0}{4d^4(n^2-1)}\frac{v-v_0}{v_0}[3+4q_0d+2(q_0d)^2]e^{-2q_0d},
\label{qvc14}
\end{equation}
\begin{equation}
w_p^{res\prime }=\frac{\hbar R^3\omega^2_0}{2d^3(n^2-1)}\frac{v-v_0}{v_0}(1+q_0d)e^{-2q_0d}]
\label{1cqvc14}
\end{equation}
where $q_0=(n^2-1)\omega_0/((v-v_0)n^2)$.
In the off-resonant region $\omega\ll\omega_0$
\[
\mathrm{Im}\alpha=R^3\frac{\omega}{\omega_0\tau},
\]
and again the dominant contributions are given by the  $p$-polarized waves (see Appendix \ref{B}):
\begin{equation}
f_{xp}^{offres\prime }\approx-\frac{5}{4\pi}\frac{\hbar R^3 v_0^2}{d^6\omega_0^2\tau}\frac{n^4}{(n^2-1)^3}\left(\frac{v-v_0}{v_0}\right)^3,
\label{qvc15}
\end{equation}
\begin{equation}
w_p^{offres\prime }\approx\frac{35}{64\pi}\frac{\hbar R^3 v_0^3}{d^6\omega_0^2\tau}\frac{n^6}{(n^2-1)^4}\left(\frac{v-v_0}{v_0}\right)^4.
\label{1qvc15}
\end{equation}
 From Eqs. (\ref{qvc14})-(\ref{1qvc15}) follows that close to the threshold velocity $w^{\prime}\ll f_x^{\prime}c$. Thus, from
Eq. (\ref{qvc17}) follows that $f_x\approx f_x^{\prime}$, which agrees with the results obtained in Ref. \cite{HenkelJPCM2015}, and $w\approx f_xv$. The  change of the kinetic energy in this limit is determined by the radiation power from a particle in the
$K$ frame, but the change of the internal energy, and consequently the change of the rest mass of the particle,
is small. The off-resonant contribution
to friction force from the frequency range $\omega \ll \omega_0$   is only important close to the threshold velocity ($(v-v_0)/v_0\ll 1$), while far from the threshold velocity the friction force
is dominated by the resonant contribution from $\omega \approx \omega_0$, as was already noted in Ref. \cite{HenkelJPCM2015}.

\begin{figure}
\includegraphics[width=1.0\textwidth]{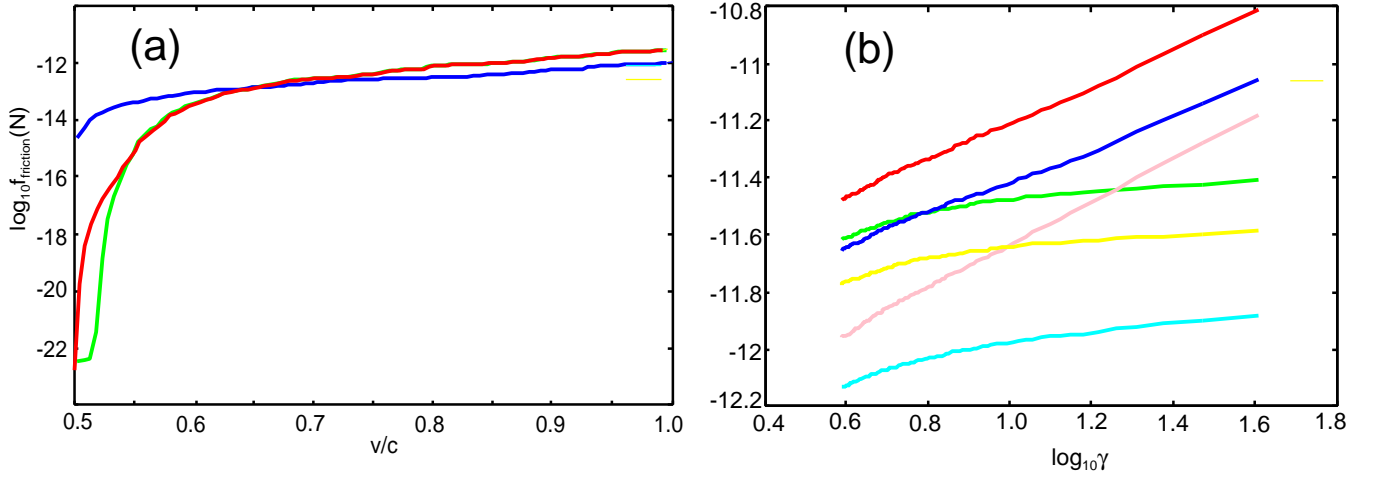}
\caption{\label{Fig.3.} The dependence of the friction force acting on a silver nanoparticle, and on the elementary charge $e$
at motion parallel to the  transparent dielectric on the relative sliding velocity. The radius of particle
$R=4$nm, the separation $d=10$nm, the refractive index  $n=2$. (a) The red and green  curves show the friction forces for a particle with losses
($\tau^{-1}=2.4\cdot 10^{14}$s$^{-1}$) and no losses ($\tau^{-1}=+0$).
The blue curve shows the friction force on a elementary charge $e$. (b) The red, blue and pink curves show the total friction force, and the contributions from the $p$- and
$s$-polarized waves, respectively, in the particle rest reference (the $K^{\prime}$ frame) in the ultra relativistic case ($\gamma \gg 1$). The green, yellow and light-blue
curves -- the same but in the \textit{lab} frame (the $K$ frame).}
\end{figure}

The friction force acting on the elementary charge $e$, due to the classical Vavilov-Cherenkov radiation, is determined  by the well-known formula
\cite{BolotovskiiUPN1962}
\begin{equation}
f_x^{clas}=-\frac{e^2}{2\beta d^2}\frac{1}{n^2-1}\left[\frac{n^2\gamma}{\sqrt{\gamma^2+n^2}}-\frac{\sqrt{n^2-1}}{\gamma}-\beta \right].
\label{cvc}
\end{equation}
In the ultra relativistic limit ($\gamma \gg 1$) Eq. (\ref{cvc}) is significantly simplified taking the form
$f_x^{clas}=-e^2/2d^2$. In Fig. \ref{Fig.3.} it is compared with the friction force due to the quantum Vavilov-Cherenkov radiation on a neutral silver nano particle with the radius $R=4$nm  and with losses
(red, $\tau^{-1}=2.4\cdot 10^{14}$s$^{-1}$)  and no losses (green,
$\tau^{-1}=+0$), the surface plasmon frequency for a particle $\omega_0=\omega_p/\sqrt{3}$, where $\omega_p$=9.01eV is the bulk plasmon frequency for silver, at  $d=10$nm.
Far from the threshold velocity these friction forces are of the same order of the magnitude. 

In the ultra relativistic limit ($\gamma \gg 1$)
\begin{equation}
\mathrm{Im}\alpha(\omega^{\prime})=-\frac{\pi}{2}\omega_0 R^3\delta(\omega^{\prime}+\omega_0)=
-\frac{\pi}{2\gamma}\frac{\omega_0 R^3}{v}\delta\left(q_x-\frac{\omega}{v}-\frac{\omega_0}{\gamma v}\right)
\end{equation}
The negative sign of the imaginary part of the particle polarizability means that a particle behaves as an object with negative absorption amplifying certain incident waves. This phenomenon is closely connected to superradiance first introduced
by Zel$^{\prime}$dovich \cite{Zeldovich}. He argued that a rotating object amplifies
certain incident waves and speculated that this would lead to
spontaneous emission when quantum mechanics is considered.The contributions to the friction force, and the radiation power, from the $s$- and $p$- polarised waves in the $K$ frame are given by (see Appendix \ref{A})

\begin{equation}
f_{sx}=-\frac{3\hbar\omega_0R^3}{2d^4(n+1)}\left[1+\left(\frac{\omega_0d}{c\gamma}\right)C\right],\,\,
p_{s}\label{sfriction}=-\frac{3\hbar v\omega_0R^3}{2d^4(n+1)}
\end{equation}

\begin{equation}
f_{px}=-\frac{3\hbar\omega_0R^3}{2d^4}\frac{n}{n+1}\left[1+\left(\frac{\omega_0d}{c\gamma}\right)C^{\prime}\right],\,\,
p_{p}=-\frac{3\hbar v\omega_0R^3}{2d^4}\frac{n}{n+1},
\label{pfriction}
\end{equation}
where
\[
C=\frac{2}{3\pi}\sqrt{\frac{n+1}{n-1}}\left[\frac{n}{\sqrt{n^2-1}}\mathrm{tanh}^{-1}\frac{\sqrt{n^2-1}}{n}-1\right],
\]
\[
C^{\prime}=\frac{2}{3\pi}n\sqrt{\frac{n+1}{n-1}}\left[\frac{n^2}{\sqrt{n^4-1}}\mathrm{tanh}^{-1}\frac{\sqrt{n^4-1}}{n^2}-
\frac{n}{\sqrt{n^2-1}}\mathrm{tanh}^{-1}\frac{\sqrt{n^2-1}}{n}\right],
\]
In the $K^{\prime}$ frame the friction force and the heat absorbed by a particle can be obtained from the corresponding quantities in the $K$ frame using the Lorenz
transformations (\ref{qvc17prime}). For example, for the contributions from the $s$-polarized waves
\begin{equation}
f_s^{\prime}=\gamma^2(f_s-\beta w_s)\approx -\frac{3\hbar\omega_0R^3}{2d^4(n+1)}\left[1+C\gamma \frac{\omega_0d}{c}\right],
\end{equation}
and
\begin{equation}
w_s^{\prime}=\gamma^2(w_s-vf_s)\approx \frac{3\hbar v\omega_0R^3}{2d^4(n+1)}C\gamma \frac{\omega_0d}{c}.
\end{equation}
Thus, contrary to the $K$, frame where the friction force and the  power of photon emission are finite, in the $K^{\prime}$ frame the friction force
and the radiation power both diverge as $\sim \gamma$. These results also can be  obtained by the direct calculations in the $K^{\prime}$ frame. The radiation power $w$ is determined mostly 
by the friction force power $f_xv$ in the $K$ frame (see Eq. (\ref{qvc17prime}). From Eq. (\ref{qvc18}) follows that the  particle heating  power also contributes significantly in this limit to  the change of the kinetic energy.

\section{Discussion}

A silver particle with $R=4$nm has mass $m_0\approx 2.68\cdot 10^{-21}$kg. Close to the threshold velocity ($v\approx v_0=c/n$) at $n=2$ it has the kinetic energy $E_K=m_0c^2(\gamma-1)\approx 2\cdot 10^2$ TeV, which is larger than the energy of proton in Large Hadron Collider  $7$ Tev. However the energy of a particle can be decreased by decreasing its radius. According to Eqs. (\ref{cvc}) and (\ref{qvc14}) the intensity of the QVC radiation exceeds intensity CVC radiation when
\begin{equation}
\hbar \omega_0 \left(\frac{R}{d}\right)^3>\frac{e^2}{2d}
\label{cond1}
\end{equation}
The point dipole approximation is valid for $d\gg R$. Assuming $d=2R$  the condition (\ref{cond1}) gives $d>4\alpha d_0$ where the fine structure constant $\alpha\approx 1/137$ and $d_0=c/\omega_0$. At typical value of $\omega_0$ in the UV range $d_0\sim 10$nm and thus $d>0.3$nm. For a particle with $R=0.4$ nm  and $d=1$nm the reduction factor for the particle mass $\sim 10^{-3}$ and the energy $\sim 0.1$ TeV. The characteristic frequency of radiation $\omega \sim v/d$. The threshold velocity $v_0$ depends on the refractive index $n$. Recently the transparent metamaterials were developed with very high reflective index in the UV range \cite{Jacobnnanotech2016}. At $n=20$ the threshold velocity $v_0\sim 10^7$m/s and at $d=1$nm the frequency of the radiation $\sim 10^{16}$s$^{-1}$ i.e. in the near UV range.  The energy of a particle in this case is  $\sim 1$GeV. Thus, in principle the QVC radiation can be detected with the present experimental setups but only in the region where relativistic effects are small. 

\section{Conclusion}
A small neutral particle moving parallel to    transparent 
dielectric plate  emits   quantum Valivov-Cherenkov (QVC) radiation  when the 
velocity exceeds a threshold velocity. This radiation is responsible for quantum friction, 
which we have studied in the particle rest reference frame  and 
in the \textit{lab} frame, using a fully relativistic theory. The friction forces in the 
particle-plate configuration   in the different reference frames were calculated from the
corresponding results in the plate-plate configuration considering one of the plates as 
sufficiently rarefied.  We have  shown that in the  realistic situation the friction force
acting on a neutral nanoparticle
due to QVC radiation
can be comparable in the magnitude with  the friction force acting on a charged particle due to the classical Vavilov-Cherenkov (CVC) radiation.  Thus, in principle QVC radiation  can be detected using the same experimental setup as for CVC radiation.    The challenges for future experiments 
are to accelerate a particle, having sufficiently large fluctuating dipole moment and  to the 
velocities close to the light velocity at  small separation  from a transparent dielectric surface. Non relativistic QVC radiation can be observed using transparent dielectric with refractive index $\sim 10$ in the near UV region.

\section{Acknowledgement}
The study was supported by the Ministry of Education and
Science of Russia under the Competitiveness Enhancement
Program of SSAU for 2013-2020, the Russian Foundation
for Basic Research  No. 16-02-00059-a, and COST Action MP1303 ÒUnderstanding and
Controlling Nano and Mesoscale Friction.Ó

\vskip 0.5cm

\appendix

\section{Particle with no losses \label{A}}

\subsection{Close to the threshold velocity}

The resonant contribution to the friction force comes from $\omega$ in the range
\[
\omega_0<\omega<\frac{v-v_0}{1-vv_0/c^2}q_x .
\]
Near resonance the particle polarazibility can be approximated by the formula
\begin{equation}
\mathrm{Im}\alpha=R^3\omega_0^2\frac{\omega\gamma}{(\omega^2-\omega_0^2)^2+\omega^2\gamma^2}\approx R^3\omega_0\frac{\pi}{2}
[\delta(\omega-\omega_0)-\delta(\omega+\omega_0)]
\end{equation}
Close to the threshold velocity ($(v-v_0)/v_0\ll 1$) and at the resonance
\begin{equation}
k_{nz}^{\prime 2}=-\frac{1}{v_0}\left[\frac{n^2}{(n^2-1)}(v-v_0)q_x-\omega_0\right]2q_x-q_y^2.
\label{c1}
\end{equation}
Introducing new variables $q_x=q_0x$, $q_y=q_0y_{max}y$ where $q_0=\omega_0(n^2-1)/((v-v_0)n^2)$, 
\[
y_{max}^2=2\frac{n^2}{n^2-1}x(x-1)\frac{v-v_0}{v_0}\ll x^2
\] 
the imaginary part of the reflection amplitudes can be written in the form
\[
\mathrm{Im}R_w^{\prime}\approx \frac{2}{x}y_{max}\sqrt{1-y^2}
\]
The $p$-wave resonant contribution to the friction force in the $K^{\prime}$ frame is given by
\[
f_x^{\prime}\approx -\frac{2^4\hbar q_0^4}{\pi^2(n^2-1)}\left(\frac{\pi}{2}R^3\omega_0\right)\frac{v-v_0}{v_0}
\int_1^{\infty}dxx^2(x-1)e^{-2q_0dx}\int_0^1dy\sqrt{1-y^2}
\]
\begin{equation}
=-\frac{\hbar R^3\omega_0}{4d^4(n^2-1)}\frac{v-v_0}{v_0}[3+4q_0d+2(q_0d)^2]e^{-2q_0d}
\label{c2}
\end{equation}
\[
w^{\prime}\approx \frac{2^4\hbar \omega_0q_0^3}{\pi^2(n^2-1)}\left(\frac{\pi}{2}R^3\omega_0\right)\frac{v-v_0}{v_0}
\int_1^{\infty}dxx(x-1)\int_0^1dy\sqrt{1-y^2}
\]
\begin{equation}
=\frac{\hbar R^3\omega^2_0}{2d^3(n^2-1)}\frac{v-v_0}{v_0}(1+q_0d)e^{-2q_0d}]
\label{c2}
\end{equation}

\subsection{Limiting case $v\rightarrow c$}

In the ultra relativistic limit ($\gamma \gg 1$) 
\begin{equation}
\mathrm{Im}\alpha(\omega^{\prime})=-\frac{\pi}{2}\omega_0 R^3\delta(\omega^{\prime}+\omega_0)= 
-\frac{\pi}{2\gamma}\frac{\omega_0 R^3}{v}\delta\left(q_x-\frac{\omega}{v}-\frac{\omega_0}{\gamma v}\right)
\end{equation}
After the integration in Eq. (\ref{qvc9}) over $q_x$
\begin{equation}
k_z^2=\left(\frac{\omega}{\gamma v}\right)^2+\frac{2\omega\omega_0}{\gamma v}+\left(\frac{\omega_0}{\gamma v}\right)^2+q_y^2
\label{1limit}
\end{equation} 
\begin{equation}
k_{nz}^2 = \left[\frac{\omega}{v}(n\beta-1)-\frac{\omega_0}{\gamma v}\right]\left[\frac{\omega}{v}(n\beta+1)+\frac{\omega_0}{\gamma v}\right]-q_y^2
\label{2limit}
\end{equation}
For $\omega < C\gamma v/d$ where $C\ll 1$: $k_z\approx q_y$,

\begin{equation}
k_{nz}^2 \approx (n^2-1)\left(\frac{\omega}{v}\right)^2-q_y^2
\label{2limit}
\end{equation}
Introducing the new variable 
\[
q_y=\frac{\omega}{v}y\sqrt{n^2-1}
\]
where $0\leq y \leq 1$,
the imaginary part of the reflection amplitudes can be written in the form
\begin{equation}
\mathrm{Im}R_s=\frac{2k_zk_{nz}}{k_z^2+k_{nz}^2}= 2y\sqrt{1-y^2},
\label{Im1}
\end{equation}
\begin{equation}
\mathrm{Im}R_p=\frac{2n^2k_zk_{nz}}{n^4k_z^2+k_{nz}^2}= \frac{2n^2y\sqrt{1-y^2}}{1+(n^4-1)y^2},
\label{Im2}
\end{equation}
and
\begin{equation}
\phi_p\approx \phi_s\approx 2\gamma^2\left(\frac{\omega}{v}\right)^2\frac{(n^2-1)^2y^4}{1+(n^2-1)y^2}.
\end{equation}
The contributions from the $s$- and $p$- polarised waves in the $K$ frame are given by
\[
f_{sx}=-\frac{4\hbar}{\pi}\omega_0R^3\int_0^1dy\frac{(n^2-1)^2y^4\sqrt{1-y^2}}{1+(n^2-1)y^2}\int_0^{\infty}d\left(\frac{\omega}{v}\right)
\left(\frac{\omega}{v}+\frac{\omega_0}{\gamma v}\right)\left(\frac{\omega}{v}\right)^2e^{-2(\omega/v)yd\sqrt{n^2-1}}
\]
\begin{equation}
=-\frac{3\hbar\omega_0R^3}{2d^4(n+1)}\left[1+\left(\frac{\omega_0d}{c\gamma}\right)C\right]
\label{sfriction}
\end{equation}
where
\[
C=\frac{2}{3\pi}\sqrt{\frac{n+1}{n-1}}\left[\frac{n}{\sqrt{n^2-1}}\mathrm{tanh}^{-1}\frac{\sqrt{n^2-1}}{n}-1\right],
\]
\[
w_{s}=-\frac{4\hbar v}{\pi}\omega_0R^3\int_0^1dy\frac{(n^2-1)^2y^4\sqrt{1-y^2}}{1+(n^2-1)y^2}\int_0^{\infty}d\left(\frac{\omega}{v}\right)
\left(\frac{\omega}{v}\right)^3e^{-2(\omega/v)yd\sqrt{n^2-1}}
\]
\begin{equation}
=-\frac{3\hbar v\omega_0R^3}{2d^4(n+1)},
\label{sradiation}
\end{equation}
\[
f_{px}=-\frac{4\hbar}{\pi}\omega_0R^3\int_0^1dy\frac{n^2(n^2-1)^2y^4\sqrt{1-y^2}}{[1+(n^2-1)y^2][1+(n^4-1)y^2]}\int_0^{\infty}d\left(\frac{\omega}{v}\right)
\left(\frac{\omega}{v}+\frac{\omega_0}{\gamma v}\right)\left(\frac{\omega}{v}\right)^2e^{-2(\omega/v)yd\sqrt{n^2-1}}
\]
\begin{equation}
=-\frac{3\hbar\omega_0R^3}{2d^4}\frac{n}{n+1}\left[1+\left(\frac{\omega_0d}{c\gamma}\right)C^{\prime}\right],
\label{pfriction}
\end{equation}
where
\[
C^{\prime}=\frac{2}{3\pi}n\sqrt{\frac{n+1}{n-1}}\left[\frac{n^2}{\sqrt{n^4-1}}\mathrm{tanh}^{-1}\frac{\sqrt{n^4-1}}{n^2}-
\frac{n}{\sqrt{n^2-1}}\mathrm{tanh}^{-1}\frac{\sqrt{n^2-1}}{n}\right],
\]
\[
w_{p}=-\frac{4\hbar}{\pi}\omega_0R^3\int_0^1dy\frac{n^2(n^2-1)^2y^4\sqrt{1-y^2}}{[1+(n^2-1)y^2][1+(n^4-1)y^2]}\int_0^{\infty}d\left(\frac{\omega}{v}\right)
\left(\frac{\omega}{v}\right)^3e^{-2(\omega/v)yd\sqrt{n^2-1}}
\]
\begin{equation}
=-\frac{3\hbar v\omega_0R^3}{2d^4}\frac{n}{n+1}.
\label{pradiation}
\end{equation}
In the deriving of the formulas (\ref{sfriction}- \ref{pradiation}) the values of the following ingrals were used
\[
\int_0^1dy\frac{\sqrt{1-y^2}}{1+(n^2-1)y^2}=\frac{\pi}{n+1}
\]
\[
\int_0^1dy\frac{y\sqrt{1-y^2}}{1+(n^2-1)y^2}=\frac{1}{n^2-1}\left[\frac{n}{\sqrt{n^2-1}}\mathrm{tanh}^{-1}\frac{\sqrt{n^2-1}}{n}-1\right]
\]
\[
\int_0^1dy\frac{\sqrt{1-y^2}}{]1+(n^2-1)y^2]1+(n^4-1)y^2]}=\frac{\pi}{n(n+1)}
\]
\[
\int_0^1dy\frac{\sqrt{1-y^2}}{]1+(n^2-1)y^2]1+(n^4-1)y^2]}=\frac{1}{n^2(n^2-1)}\left[\frac{n^2}{\sqrt{n^4-1}}\mathrm{tanh}^{-1}\frac{\sqrt{n^4-1}}{n^2}-
\frac{n}{\sqrt{n^2-1}}\mathrm{tanh}^{-1}\frac{\sqrt{n^2-1}}{n}\right],
\]
In the $K^{\prime}$ frame the friction force and the heat absorbed by a particle can be obtained from the corresponding quantities in the $K$ frame using the Lorenz 
transformations (\ref{qvc17prime}). For example for the contributions from the $s$-polarized waves 
\begin{equation}
f_s^{\prime}=\gamma^2(f_s-\beta w_s)\approx -\frac{3\hbar\omega_0R^3}{2d^4(n+1)}\left[1+C\gamma \frac{\omega_0d}{c}\right],
\end{equation}
and
\begin{equation} 
w_s^{\prime}=\gamma^2(w_s-vf_s)\approx \frac{3\hbar v\omega_0R^3}{2d^4(n+1)}C\gamma \frac{\omega_0d}{c}.
\end{equation}
Thus, contrary to the $K$ frame where the friction force and the  power of photon emission are finite, in the $K^{\prime}$ frame the friction force and
the radiation power both diverge as $(1-\beta)^{-1/2}$. These results can be confirmed by the direct calculations in the $K^{\prime}$
where
\begin{equation}
\mathrm{Im}\alpha(\omega)=-\frac{\pi}{2}\omega_0 R^3\delta(\omega-\omega_0)
\end{equation}
After the integration in Eq. (\ref{qvc11}) over $\omega$
\begin{equation}
k_z^2=q_x^2+q_y^2-\left(\frac{\omega_0}{c}\right)^2
\label{2limit}
\end{equation} 
\[
k_{nz}^{\prime 2} = \gamma^2\left[n^2\beta^2\left(q_x-\frac{\omega_0}{v}\right)^2-\left(q_x-\frac{\omega_0}{c}\right)^2\right]-q_y^2
\label{2limit}
\]
\begin{equation}
\approx \gamma^2(n^2-1)\left(q_x-\frac{\omega_0}{v}\right)^2-q_y^2
\label{2limit}
\end{equation}
Introducing the new variable 
\[
q_y=\gamma\sqrt{n^2-1}\left(q_x-\frac{\omega_0}{v}\right)y
\]
where $0\leq y \leq 1$ and $q_x\geq \omega_0/v$, the same equations (\ref{Im1}) and (\ref{Im2}) are obtained for $\textrm{Im}R_{p(s)}$ and 
\begin{equation}
\phi_p\approx \phi_s\approx 2\gamma^4\left(q_x-\frac{\omega_0}{v}\right)^2\frac{(n^2-1)^2y^4}{1+(n^2-1)y^2}.
\end{equation}
The contributions from the $s$- and $p$- polarised waves in the $K^{\prime}$ frame are given by
\[
f_{sx}^{\prime}=-\frac{4\hbar}{\pi}\omega_0R^3\gamma^4\int_0^1dy\frac{(n^2-1)^2y^4\sqrt{1-y^2}}{1+(n^2-1)y^2}\int_0^{\infty}dq_x
q_x\left(q_x-\frac{\omega_0}{ v}\right)^2e^{-2(q_x-\omega_0/v)yd\gamma\sqrt{n^2-1}}
\]
\begin{equation}
=-\frac{3\hbar\omega_0R^3}{2d^4(n+1)}\left[1+C\gamma \frac{\omega_0d}{c}\right],
\end{equation}
\[
w_{s}^{\prime}=\frac{4\hbar}{\pi}\omega_0^2R^3\gamma^4\int_0^1dy\frac{(n^2-1)^2y^4\sqrt{1-y^2}}{1+(n^2-1)y^2}\int_0^{\infty}dq_x
\left(q_x-\frac{\omega_0}{ v}\right)^2e^{-2(q_x-\omega_0/v)yd\gamma\sqrt{n^2-1}}
\]
\begin{equation}
=\frac{3\hbar v\omega_0R^3}{2d^4(n+1)}C\gamma \frac{\omega_0d}{c}.
\end{equation}

\section{The off-resonant contribution for a particle with losses close  to the threshold velocity \label{B}}

The off-resonant contribution to the friction force comes from $\omega$ in the range
\[
0<\omega<\frac{v-v_0}{1-vv_0/c^2}q_x\ll \omega_0.
\]
In this frequency range the low-frequency approximation for the particle polarisability can be used
\[
\mathrm{Im}\alpha=R^3\omega_0^2\frac{\omega\gamma}{(\omega^2-\omega_0^2)^2+\omega^2\gamma^2}\approx R^3\frac{\omega\gamma}{\omega_0^2}
\]
Close to the threshold velocity ($(v-v_0)/v_0\ll 1$) $\omega$ is small and to lowest order in $(v-v_0)/v_0\ll 1$
\[
k_{nz}^{\prime 2}=\left(\frac{\omega^{\prime}}{v_0}\right)^2-q^{\prime 2}
\]
\[
=\gamma^2 \frac{[q_x(v-v_0)-\omega(1-vv_0/c^2)][q_x(v+v_0)-\omega(1+vv_0/c^2]}{v^2_0}-q_y^2
\]
\begin{equation}
=\frac{1}{v_0}\left[\frac{n^2}{(n^2-1)}(v-v_0)q_x-\omega\right]2q_x-q_y^2.
\label{b1}
\end{equation}
The integration over $q_y$ is restricted by the range $0<q_y<q_xy_{max}$, where
\[
y_{max}^2=2\frac{n^2}{n^2-1}\frac{v-v_0}{v_0}\ll 1.
\]

Introducing new variables $q_y=q_xy_{max}y$, 
\[
\omega=q_xv_0y_{max}^2\frac{1-y^2}{2}z
\]
the imaginary part of the reflection amplitudes can be written in the form
\[
\mathrm{Im}R_s^{\prime}=\frac{2k_zk_{zn}^{\prime}}{k_z^2+k_{zn}^{\prime 2}}\approx\frac{2k_{zn}^{\prime}}{k_z}\approx \frac{2}{q_x}\sqrt{\frac{1}{v_0}\left[\frac{n^2}{(n^2-1)}(v-v_0)q_x-\omega\right]2q_x-q_y^2}
\]
\begin{equation}
=2y_{max}\sqrt{1-y^2}\sqrt{1-z} \sim \sqrt{\frac{v-v_0}{v_0}},
\label{a4}
\end{equation}
and
$\mathrm{Im}R_s^{\prime}=\mathrm{Im}R_s^{\prime}/n^2$,
\[
\phi_s^{\prime}\approx 2q_x^2,\,\, \,\phi_s^{\prime}\sim q_x^2y_{max}^2\ll q_x^2.
\]
To lowest order in $(v-v_0)/v_0$ the friction force is determined only by the contribution from the $p$-polarized waves which is given by
\[
f_x^{\prime}\approx  \frac{2\hbar y_{max}^6v_0^2R^3}{n^2\pi^2}\int_0^{\infty}dq_xq_x^5e^{-2q_xd}\int_0^1dy(1-y^2)^{5/2}
\int_0^1dzz\sqrt{1-z}
\]
\begin{equation}
=-\frac{5}{4\pi}\frac{\hbar R^3 v_0^2}{d^6\omega_0^2\tau}\frac{n^4}{(n^2-1)^3}\left(\frac{v-v_0}{v_0}\right)^3,
\end{equation}
and the heat absorbed by the particle in the $K^{\prime}$ frame is given by
\[
w^{\prime}\approx  \frac{\hbar y_{max}^8v_0^3R^3}{n^2\pi^2}\int_0^{\infty}dq_xq_x^5e^{-2q_xd}\int_0^1dy(1-y^2)^{7/2}
\int_0^1dzz\sqrt{1-z}
\]
\begin{equation}
=\frac{35}{64\pi}\frac{\hbar R^3 v_0^3}{d^6\omega_0^2\tau}\frac{n^6}{(n^2-1)^4}\left(\frac{v-v_0}{v_0}\right)^4,
\end{equation}

\end{document}